# RECONSTRUCTION OF HYDRAULIC DATA BY MACHINE LEARNING

Corentin J. Lapeyre, Nicolas Cazard, Pamphile T. Roy, Sophie Ricci
CECI, CERFACS/CNRS, 42 avenue Gaspard Coriolis, 31057 Toulouse Cedex 01, France
Phone: 0561193128, lapeyre@cerfacs.fr

Fabrice Zaoui
EDF/LNHE, 6 quai Watier, 78401 Chatou, France
Phone: 0130877657, fabrice.zaoui@edf.fr

**ABSTRACT**

*Numerical simulation models associated with hydraulic engineering take a wide array of data into account to produce predictions: rainfall contribution to the drainage basin (characterized by soil nature, infiltration capacity and moisture), current water height in the river, topography, nature and geometry of the river bed, etc. This data is tainted with uncertainties related to an imperfect knowledge of the field, measurement errors on the physical parameters calibrating the equations of physics, an approximation of the latter, etc. These uncertainties can lead the model to overestimate or underestimate the flow and height of the river. Moreover, complex assimilation models often require numerous evaluations of physical solvers to evaluate these uncertainties, limiting their use for some real-time operational applications.*

*For problems with high uncertainty and vast amounts of measurements available such as hydraulics, a new emerging paradigm has been gaining traction in recent years, namely the data-driven approach. Based mostly on machine learning techniques, these optimization techniques aim to build fast surrogate models entirely inferred from the data. Indeed, a large variety of function classes are available today in this context, and can be rapidly tested to find those who best match the underlying trends in the data. In this approach, these trends are therefore not hand-designed by physicists, but selected based on performance on a given dataset.*

*In this study, we explore the possibility of building a predictor for river height at an observation point based on drainage basin time series data. An array of data-driven techniques is assessed for this task, including statistical models, machine learning techniques and deep neural network approaches. These are assessed on several metrics, offering an overview of the possibilities related to hydraulic time-series. An important finding is that for the same hydraulic quantity, the best predictors vary depending on whether the data is produced using a physical model or real observations.*

**INTRODUCTION**

Machine Learning (ML) is a highly successful strategy in many digital industries, where it has demonstrated capabilities far superior to previous approaches in key sectors such as image or text analysis. Applications in scientific computing, on the other hand, are still limited. Nevertheless, the performance of ML, and specifically of a class of techniques called Deep Learning (DL) suggests that many new applications will emerge in the coming years.

Hydraulic engineering aims to explain and predict the water elevation and discharge of rivers. The associated numerical simulation models take into account rainfall over the catchment area (characterized by the nature, infiltration capacity and soil moisture), often translated into input discharge as well as bathymetry/topography and friction for the river bed and flood plain. These data are subject to uncertainties related to imperfect knowledge of the digital elevation model, measurement errors in the data used to calibrate the numerical model, simplification of the physics equations in the numerical model. These uncertainties may lead the model to overestimate or underestimate the discharge and water lever. The characterization of river hydrodynamics and its prediction with increasing lead time (especially beyond the transfer time of the hydraulic network) could be considerably enriched by taking into account the different sources of model uncertainty. In this perspective, stochastic methods are commonly used in the field of computer experiments to appropriately analyze the sensitivity of the model's outputs to its inputs. However, Monte-Carlo derived strategies require a large number of model evaluations, associated to a significant computational cost, especially in real-time mode. Thus, it may be appropriate to replace the direct solver with a parsimonious approximate model - a substitution model [14, 15].

We propose here to build a model to infer the hydraulic state at a given location in the river based on a time series of boundary conditions (hydrological input discharge upstream of the river and free surface elevation downstream of the river). A range of learning tools are explored such as kriking, gradient boosting and neural networks. Their results are assessed with respect to a linear regression baseline, on the Garonne river benchmark between Tonneins and La Réole (TLR) [1]. The input data are the discharge time series and water level at Tonneins and La Réole respectively ($q_{Ton}$ and $h_{Lar}$), as well as the water level at Marmande (a city prone to flooding). The study is carried out over two learning tasks, first using simulated water level computed with the Mascaret solver [4], then using observed water level at Marmande.

The structure of the paper is as follows: Section 1 describes the materials and methods for the study. It includes the description of the hydraulic numerical solver for the 1D SWE and the Garonne benchmark. The learning algorithms are described in Section 2. Results are presented in Section 3. Conclusion and perspectives are given in Section 4.

# 1 MATERIAL: NUMERICAL SOLVER AND LEARNING DATA

## 1.1 The hydraulic numerical solver for the 1D SWE

A 1D-with storage areas solver for the SWE was applied over the Garonne TLR reach. The main features of this solver are recalled here, the reader should refer to [1] for further description. The 1D SWE are written in terms of hydraulic section $A(x,t)$ and discharge $q(x,t)$ by the continuity equation (Equation 1) and the momentum equation (Equation 2).

$$\frac{\partial A}{\partial t} + \frac{\partial q}{\partial x} = Q_L \quad (1)$$

$$\frac{\partial q}{\partial t} + \frac{\partial q^2/A}{\partial x} + gA(S_f + \frac{\partial z}{\partial x}) = C_L Q_L v \quad (2)$$

$x$ is the abscissae along the hydraulic axis of the reach, $t \in [0,T]$ where, $q(x,t)$ is the local discharge, $z(x,t)$ is the water surface elevation (WSE), $Q_L(x,t)$ is the lateral discharge, $C_L(x,t)$ is the lateral discharge coefficient, $v(x,t) = \frac{q}{A}$ is the mean velocity and $S_f$ is the friction term that depends on the Strickler coefficient $K_s(x)$ :

$$S_f = \frac{q^2}{K_S^2 A^2 R^{4/3}} \quad (3)$$

The initial condition for SWE is:

$$z(x,0) = z_0(x) \text{ and } q(x,0) = q_0(x) \quad (4)$$

In a 1D model [5], the stream channel is described by a hydraulic axis corresponding to the main direction of the flow. The river channel is represented by a series of cross-sections (or profiles). Each section, identified by its curvilinear abscissa, can be divided in three zones: the main channel, the overbank flow channel (i.e. the floodplain inundated during high flow) and the storage area (low or nil current). These three zones constitute a compound channel with storage area. The numerical solvers for the 1D/1.5 SWE used in the present is Mascaret [4], an open-source software developed by EDF R&D and CEREMA (http/www.opentelemac.org).

The area chosen for this comparison extends over a 50 km reach of the Garonne river between Tonneins, downstream of the confluence with the river Lot, and La Réole (Figure 1-a). This part of the valley was equipped in the 19[th] century with infrastructure to protect against floods on the Garonne which had heavily impacted local residents, particularly since the historic flood of 1875. A system of longitudinal dykes and weirs was progressively constructed after that flood event to protect the floodplains, organize submersion and flood retention areas. Protections on the river Garonne form a system of successive storage areas for the flood plain beyond the dikes. The TLR reach is thus well adapted to 1D storage area modeling. It is a similar configuration to that of other managed rivers such as the Rhône and the Loire. The bathymetry for the TLR reach is shown in Figure 1-b along with the 3 areas where the friction coefficient homogeneous.

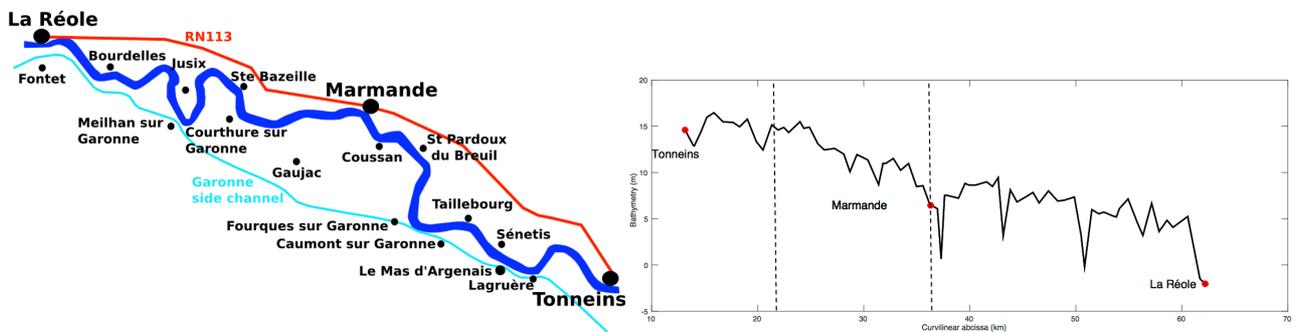

**Figures 1**: a- Garonne river reach TLR between Tonneins and La Réole, 1D reach with storage areas, b - Bathymetry of the TLR reach and friction areas separated by vertical dashed lines.

### 1.2 Training and testing datasets

A learning task must be set up in a specific manner: one must clearly define the authorized input variables, and matching output to be *predicted*. Each learning algorithm is classically referred to as a *model*, and in order to avoid confusion, Mascaret will be referred to as the *physical model*. We choose here to define a task that can be achieved by the learned models and the physical one: estimate the water level at Marmande, given the upstream discharge at Tonneins and the downstream water level at La Réole (respectively mass flow rate at Tonneins $q_{Ton}$ and outlet water lever at La Réole $h_{Lar}$). The input data for the learning task are the boundary conditions time series ($q_{Ton}, h_{Lar}$), and the target data for the learning task is the water level time series at Marmande $h_{mar}$. Additionally to these inputs, the physical model also requires the geometry of the river. In practice, this corresponds to building a surrogate model for the behavior of the time-series of water level at Marmande given the two input time-series $h_{mar} = f(q_{Ton}, h_{Lar})$. Following a classical machine learning best-practice, the data collected hourly between 1996 and 2004 is split in two: years 1996 to 2002 are used for training, and years 2003–2004 are set aside, and called the *testing set*. $h_{mar}$ for the testing set is never observed during training, and is used *a posteriori* to check the capacity of the models to generalize to previously unseen years. Two separate tasks were setup for this same period:
- in Task 1, the output $h_{mar}$ of the simulated with the physical model every 24 hours is used as a target,
- in Task 2, the hourly in-situ observations $h_{mar}$ at Marmande are used.

The Mascaret simulations were run on the raw boundary conditions data, yet the gap in the time series are automatically interpolated linearly by Mascaret when used. For a fair comparison between physical and learning models, the training in Task 1 and Task 2 was therefore performed on an hourly linear interpolation of the ($q_{Ton}, h_{Lar}$) data. In the case of Task 2, and additional cleaning of the input testing data was performed, were some obviously bad points were removed (especially at La Réole), in order to have the most objective comparison between techniques. Note that no cleaning was performed on the training data, as in any realistic case the training data is too big to fully clean, and learning algorithms must be robust to a certain level of errors in the training data.

|  |  | Train (1996 - 2002) | | | Test (2003 - 2004) | | |
|---|---|---|---|---|---|---|---|
|  |  | $q_{Ton}$ | $h_{Lar}$ | $h_{mar}$ | $q_{Ton}$ | $h_{Lar}$ | $h_{mar}$ |
| **Task 1** $h_{mar}$ from Mascaret | dt | hourly | hourly | daily | hourly | hourly | daily |
|  | interp | yes | yes | no | yes | yes | no |
|  | cleaning | no | no | no | no | no | no |
| **Task 2** $h_{mar}$ from obs | dt | hourly | hourly | hourly | hourly | hourly | hourly |
|  | interp | yes | yes | yes | yes | yes | yes |
|  | cleaning | no | no | no | no | yes | yes |

Table 1: Setups for the two tasks.

Figure 2 gives a view of the target data $h_{mar}$ that are sought to be predicted in task 2.

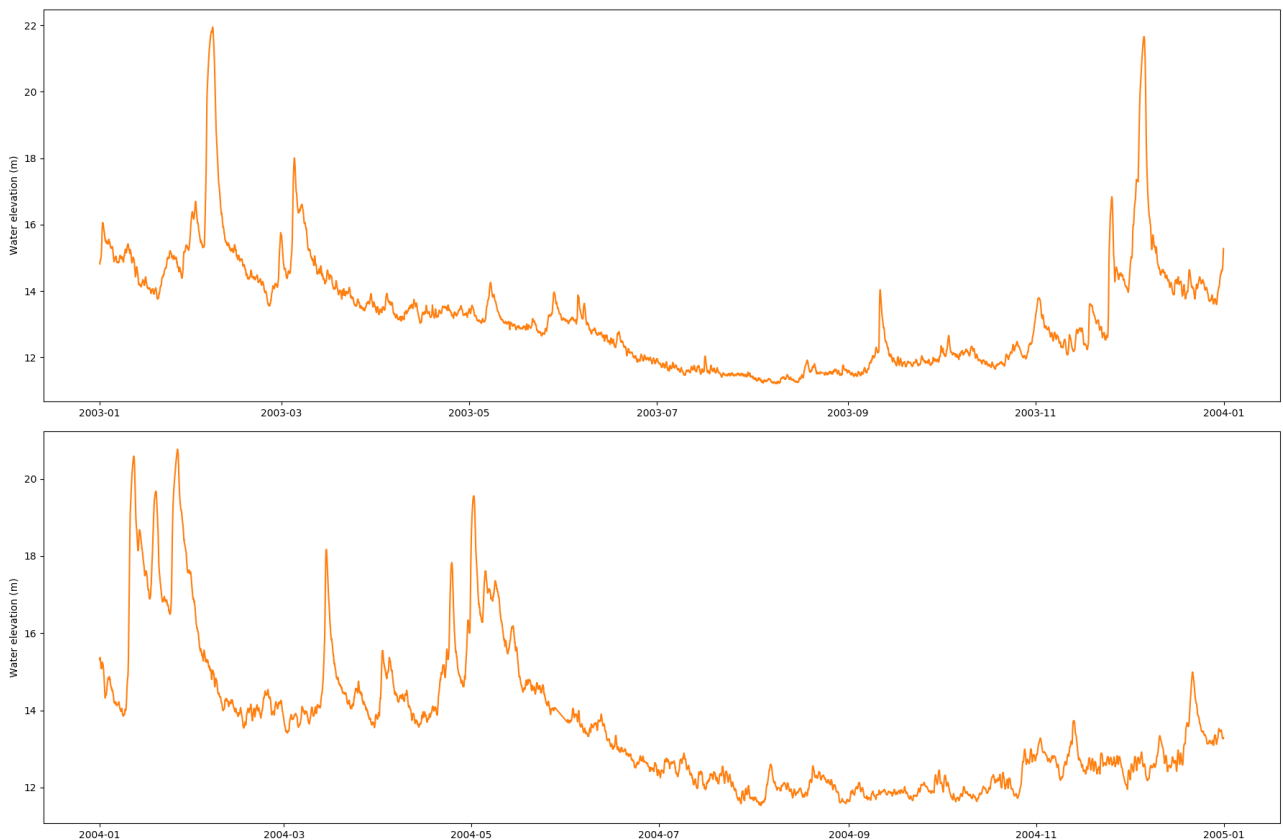

**Figures 2**: Temporal series of the water elevation at Marmande for years 2003 and 2004.

## 2. MACHINE LEARNING METHODOLOGY

### 2.1 Learning strategy

The input data is noted $x = (x_1, \ldots, x_n)$ and the output is noted $y = (y_1, \ldots, y_n)$. The learning strategy aims at approximating the true relation between $x$ and $y$ that reads : $y = f(x) + \varepsilon$, with $f$ a model and $\varepsilon$ the noise with a mean of 0. The learning model is denoted $\hat{f}$ and its MSE on a given set of observations $(x_1, y_1), \ldots, (x_n, y_n)$ reads:

$$MSE = \sum_{i=1}^{n}(y_i - \hat{f}(x_i))^2.$$

Using data at Tonneins and La Réole at time *t*, a simple correlation (*i.e.* without learning a complex model from the data) already performs well with a *root mean square error* (RMSE) of 18.5 cm on the test set for Task 1. This compares to the annual variability of the water level of the order of several meters and an aimed precision of the water level estimation of less than 10 cm. The success of the learning model on the training task should thus be assessed with respect to a *baseline*, chosen as the regression strategy. Learning algorithms are therefore evaluated on their capacity to increase the accuracy above the score of the baseline, using a metric referred to as the *fraction of explained residual* (FER):

$$FER = 1 - \frac{MSE_{model}}{MSE_{reg}},$$

with $MSE_{reg}$ computed with $\hat{f} = f_{reg}$ and $MSE_{model}$ computed with $\hat{f} = f_{model}$. This score has a simple interpretation: 1 would mean that the data is perfectly predicted with the learning model; 0 means that the prediction is no better than a linear regression. Some training algorithms can yield *negative* scores, performing worse than the regression with no training, and should be eliminated.

### 2.2 Learning algorithms

#### 2.2.1 Gaussian Process Regression

Gaussian Process (GP) regression is a classical machine learning method [10]. It is also known as *Kriging* in the field of geostatistics [11]. This method allows to construct an interpolator by considering that the value at one point is conditioned by the value of its neighbours. A correlation matrix is used to define these relationship.

A GP is a collection of random variables which have a joint Gaussian distribution. A Gaussian Process is described by its mean $\mu(x)$ and covariance $k(x, x')$—where $x, x'$ are different sets of inputs

$$Y(x) \sim GP(\mu(x), k(x, x')), \text{with}$$
$$m(x) = E[Y(x)],$$
$$k(x, x') = E[(Y(x) - \mu(x))(Y(x') - \mu(x'))].$$

Here the covariance function $k$ (or kernel) is chosen as a Matérn-3/2

$$K = k(x, x') = \sigma_x^2 \left(1 + \frac{\sqrt{3} \, \|x - x'\|^2}{l}\right) exp\left(-\frac{\sqrt{3} \, \|x - x'\|^2}{l}\right),$$

where $l$ is a correlation length scale between one sample and another (boundary condition time series), and $\sigma_x$ is the variance of the output signal (water level at Marmande). Other kernel functions can be considered, such as a decreasing exponential one or a squared exponential one—with their associated hyperparameters. The choice of the kernel is still an open problem and can be mitigated using the available information on the problem. The Matérn kernel leads to stable results and allows for more non-linearities than a classical squared exponential kernel.

Then the GP model consists of a regression providing an interpolation $M_{gp}$ for a new set of input parameters $x^*$:

$$M_{gp}(x^*) = \underline{Y}(x^*) = \sum_{i=1}^{N_s} \alpha_i k(x^i, x^*), \quad \text{with}$$
$$\alpha = (K + \sigma_n^2 \, I)^{-1} Y$$

where $\underline{Y}$ is the mean realization, $x^i$ the i-th set of parameters, $Y$ the sample matrix of water level at Marmande and $\sigma_n$ is the nugget effect that prevent ill-conditioning issues for the matrix $K$. Indeed, it is the mean realization of the conditioned process considering an artificial noisy observation which gives the prediction.

A key advantage of this predictor is that it provides an inference about its prediction variance

$$v[Y(x^*)] = k(x^*, x^*) - k(x^*)^T (K + \sigma_n^2 I)^{-1} k(x^*).$$

The learning phase of the GP consists in selecting $l$, $\sigma_n$ and $\sigma_x$ so that $Y$ passes through or close to the dataset points. The perfect interpolant property of the GP is relaxed by changing the diagonal of the correlation matrix. This is used to take into account some noise in the data. This adds another hyperparameter to fit. The hyperparameters are optimized by maximizing the log likelihood applied to the data set $Y$ using a basin hopping technique [12].

The length scale $l$ is optimized per dimension of the input parameter space. Using a long time periode to predict the water level at Marmande, this parameter space can be huge. Considering 24h of hourly data to predict the water level, the input parameter space size would be 48 (24 elements for mass flow rate at Tonneins and 24 elements for the water elevation at La Réole). This, resulting in a 48 dimension optimization problem which can be challenging.

This issue can be mitigated using data reduction technique such as PCA [13]. The input parameter space was separated into two matrices $x_Q$ and $x_H$ of size $(n, n_{dim})$ with $n$ the number of samples and $n_{dim}$ the number of hour to learn from. These matrices represent the mass flow rate at Tonneins and the water elevation at La Réole, respectively. Thus, each matrix can be decomposed as

$$x = U\Lambda V^T = \sum_{k=1}^{r} \lambda u_k v_k^T,$$

where $U \in \mathbb{M}_{n_{dim}}(\mathbb{R})$ is an orthogonal matrix diagonalizing $xx^T$ ($u_k$, the $k$th column of $U$, is a left singular vector of $x$), where $V \in \mathbb{M}_n(\mathbb{R})$ is an orthogonal matrix diagonalizing $x^T x$ ($v_k$, the $k$th column of $V$, is a right singular vector of $x$), and where $\Lambda \in \mathbb{M}_{n_{dim},n}(\mathbb{R})$ is a rectangular diagonal matrix including $r = \min(n_{dim}, n)$ singular values on its diagonal. The singular values $\{\lambda_k\}_{1 \leq k \leq r}$ are the square roots of the eigenvalues.

Using this decomposition, the input parameter space is reduced to a few random variables. Hence, the model is fitted onto this modal parameter space and to predict a new sample, it has to be converted into this reduced space. A similar approach was used with success to predict the water elevation along the Garonne river in [16].

The parametrization of the algorithm itself (number of samples, number of modes to keep and number of hours to use) has been optimized using a bayesian optimization [17].

*2.2.2 Gradient boosted trees*

The concept of gradient boosting [6] is popular in the machine learning community, and often appears in the winning submissions of classification and regression competitions. It expands on the concept of using decision trees by stacking them to refine the predictions. Indeed, suppose the task is defined as a list of data tuples $(x_1, y_1), \ldots, (x_n, y_n)$. First, a regression tree is trained on the task, yielding a function $F$. $F(x_i)$ yields $y_i$ with a given accuracy, but could be improved. A new model $h$ is introduced to reduce the error, such that $F(x_i) + h(x_i) = y_i$. The new task is to train $h$ on the data points $(x_1, y_1 - F(x_1)), \ldots, (x_n, y_n - F(x_n))$. This iterative process usually goes on for a large number of steps (100 by default in the scikit-learn library used here), and when applied using only regression trees is referred to as gradient boosted trees (GBT).

*2.2.3 Multi-Layer Perceptron*

Deep learning methods have reached a high popularity in some machine learning tasks. While these models can be very complex, training state-of-the-art neural networks also requires notoriously large amounts of data. In the context of this study, a very simple neural network called the *multi-layer perceptron* (MLP) is used. It is a simple assembly of basic feedforward "neurons" [7]. A neuron $k$ is a function that takes all the inputs from the previous layer $(x_1, \ldots, x_n)$, performs a weighted sum with weights $(w_{k,1}, \ldots, w_{k,n})$ plus a bias $b_k$, and yields the result $y_k$ through an activation function $\sigma$:

$$y_k = \sigma\left(\sum_{i=1}^{n} w_{k,i} x_i + b_k\right).$$

In the context of this study, $\sigma$ is the so-called *rectified linear unit* (ReLU) function, or simply $\sigma(x) = max(0, x)$. One such layer is known as a *fully connected layer*, and the *trainable weights* are $w_{k,1}, \ldots, w_{k,n}$ and $b_k$.

Originally, single-layer networks were used, and referred to as *perceptrons* [8], but were later expanded upon by using several layers feeding-into each other [9]. This latter form was branded the *multi-layer perceptron*, and it yielded many interesting results in the 1980's. Our model is comprised of 7 fully connected layers with 100 neurons each, and a last one with 1 neuron that outputs the target value.

*2.2.4 Convolutional Neural Network*

A Convolutional Neural Network (CNN) is a class of neural networks which responsible for many state-of-the art performances, *e.g.* in computer vision and natural language processing. Compared to an MLP, a CNN is composed of convolutional layers which make use of a set of learnable filters. At each layer, a series of filters is convolved along the temporal dimension to detect specific features or patterns. The result, called an *activation map*, is then passed to the next layer in the CNN. Our model is comprised of 3 convolutional layers with 128 filters each, and 3 fully connected layers with 40, 20 and 1 neuron.

**3. RESULTS**

The 4 techniques described in Sec. 2.2 have been used to train models to perform the two tasks described in Tab. 1, in addition to the linear regression baseline.

Results on the FER metric, as well as RMSE are shown in Fig. 3.

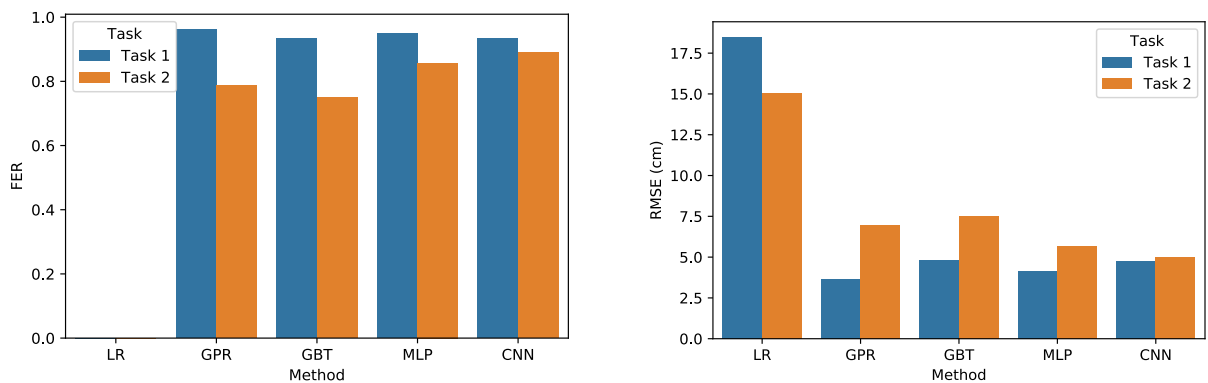

**Figures 3**: FER (left) and RMSE (in cm, right) for each learning method.
LR: Linear Regression; GPR: Gaussian Process Regression; GTB: Gradient Tree Boosting; MLP: Multi-Layer Perceptron; CNN: Convolutional Neural Network.

Several interesting observations arise from these graphs:
- The learning techniques selected here have been trained to levels of accuracy significantly superior to linear regression, as a minimum of 70% of the residual variance is explained as shown in the FER histogram in Fig. 3.
- The RMSE on task 1 is below 5 cm for all techniques. The best technique is GPR with 3.6 cm.
- The RMSE on task 2 is higher, and only the CNN reaches a precision of 5 cm. It seems the in-situ data is harder to predict than the data simulated from the physical model.
- The best technique on task 1 (GPR) is almost twice less accurate on task 2. CNN however is almost insensitive to the swap between the tasks.

This suggests an interesting, twofold observation:
- GPR, well known as kriging in the hydraulics community as a surrogate technique, indeed performs very well on data stemming from a physical model. However, even through numerous optimization loops, we were unable to obtain satisfying results compared to the MLP or CNN.
- Neural networks proved more robust to the swap to realistic data, the most extreme case being the CNN for which the RMSE only went from 4.7 cm (task 1) to 5.0 cm (task 2).

While all learning models are trained based on a loss function based on a mean square error metric, the maximum error metric is not taken into account in anyway in the learning process. Yet, in the context hydraulic and flood surveillance, the prediction of extreme events is of major interest and assessing the quality of a learning model on metrics based on mean square errors is not sufficient. Indeed, while there are many observations of "normal" conditions to train on, it is important that the learning model also accurately captures the occurrences of extreme events, even though it was trained on only a handful of them. Thus, the maximum error was computed as a diagnostic and shown in Figure 4 for all the models. The linear regression upwards of 1.5 m of maximum error, and the models all improve on this. However, the maximum error remains of the order of 50 cm for all learning models, on both tasks, which is significant in the perspective of informing decision support systems. Further study should investigate the capacity of predicting extreme events depending of the number of occurence in the training set. This type of failure is illustrated in Figure 5-a over the 01-10/12/2003 period. The GTB model is unable to predict a flood peak beyond the maximum value observed in the training set, as opposed to other methods that are able to extrapolate the prediction, even though with critiqueable results as enhanced in Figure 5-b. All model succeed in predicting the flood rise while they tend to overestimate the water level during the flood decrease, this remark is coherent with a common fact in hydrology stating that the dynamic of flood decrease is complex and difficult to simulate.

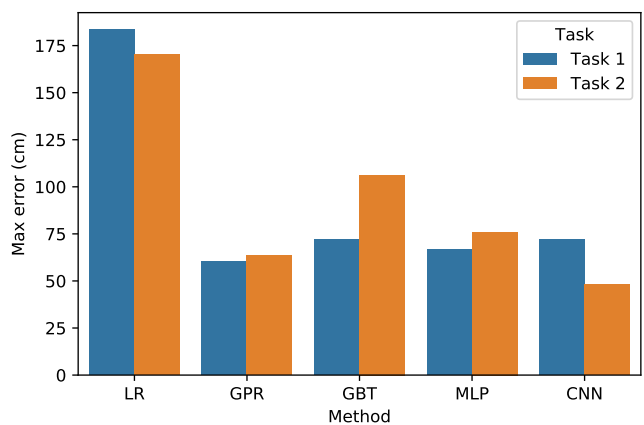

**Figure 4**: Maximum error of all trained models on the full test set.

The difference between the target and the predicted water level time series for the test period are shown in Figure 6 for GPR, GTB and CNN. For clarity purpose, the MLP model was left aside as its results are close to that of CNN, the most advanced solution was prefered. The water elevation is also represented. As expected, this visualization highlights that the worst results obtained with the learning model occur for flood peaks and dry periods that are poorly represented with in the training data set. The Probability Density Function (PDF) of the predicted water level errors with respect to the target for the test set are shown in Figure 7. All three PDF are well approximated by gaussian functions. It appears that the GPR has a positive bias of about 50 cm. Both CNN and GTB are unbiased, with a larger variance for the GTB.

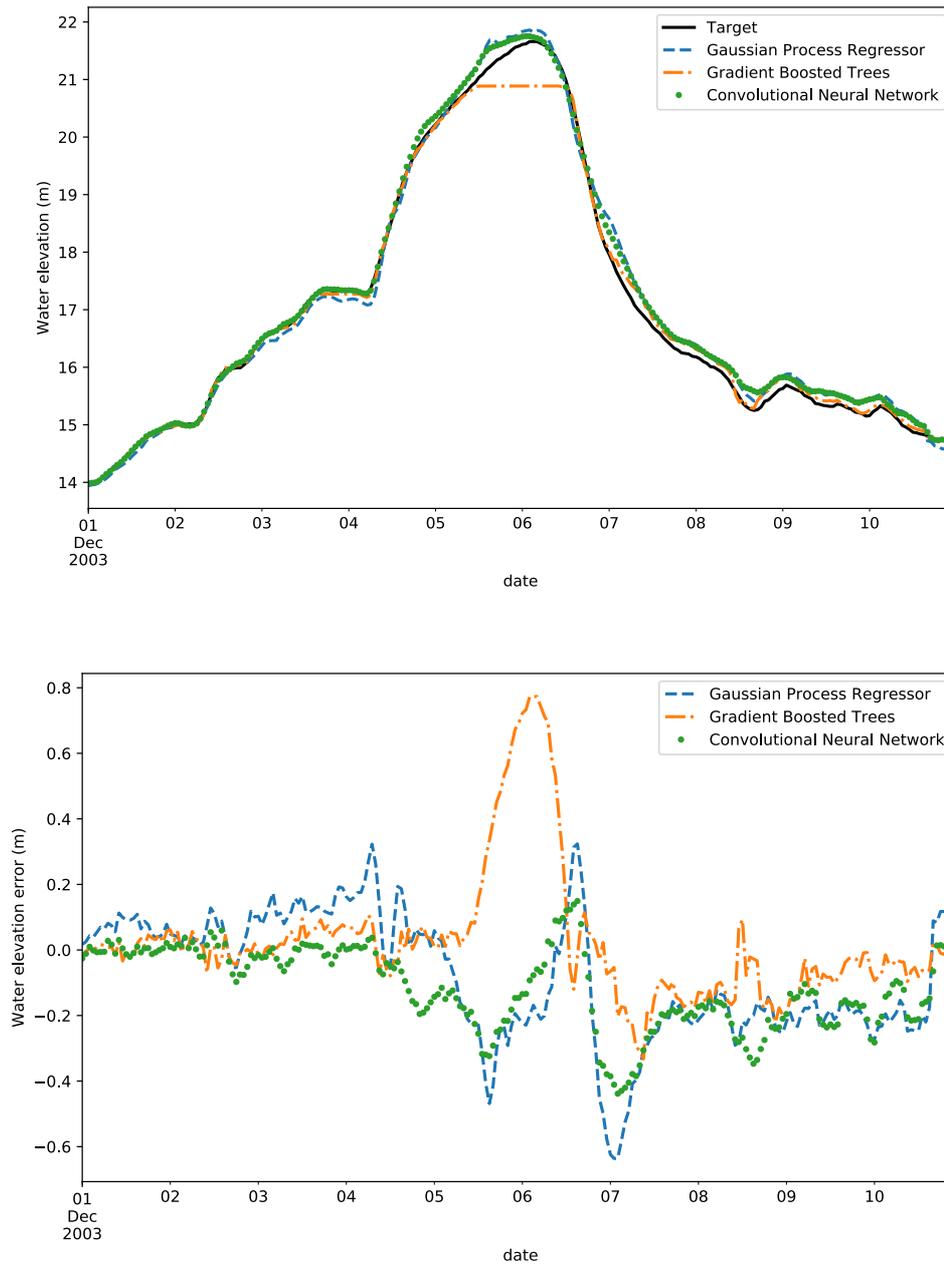

**Figure 5**: a - Predicted water level for the tests set, b- Difference between the target (black) and the predicted water level and for a 10-day event in December 2003 (date in days).

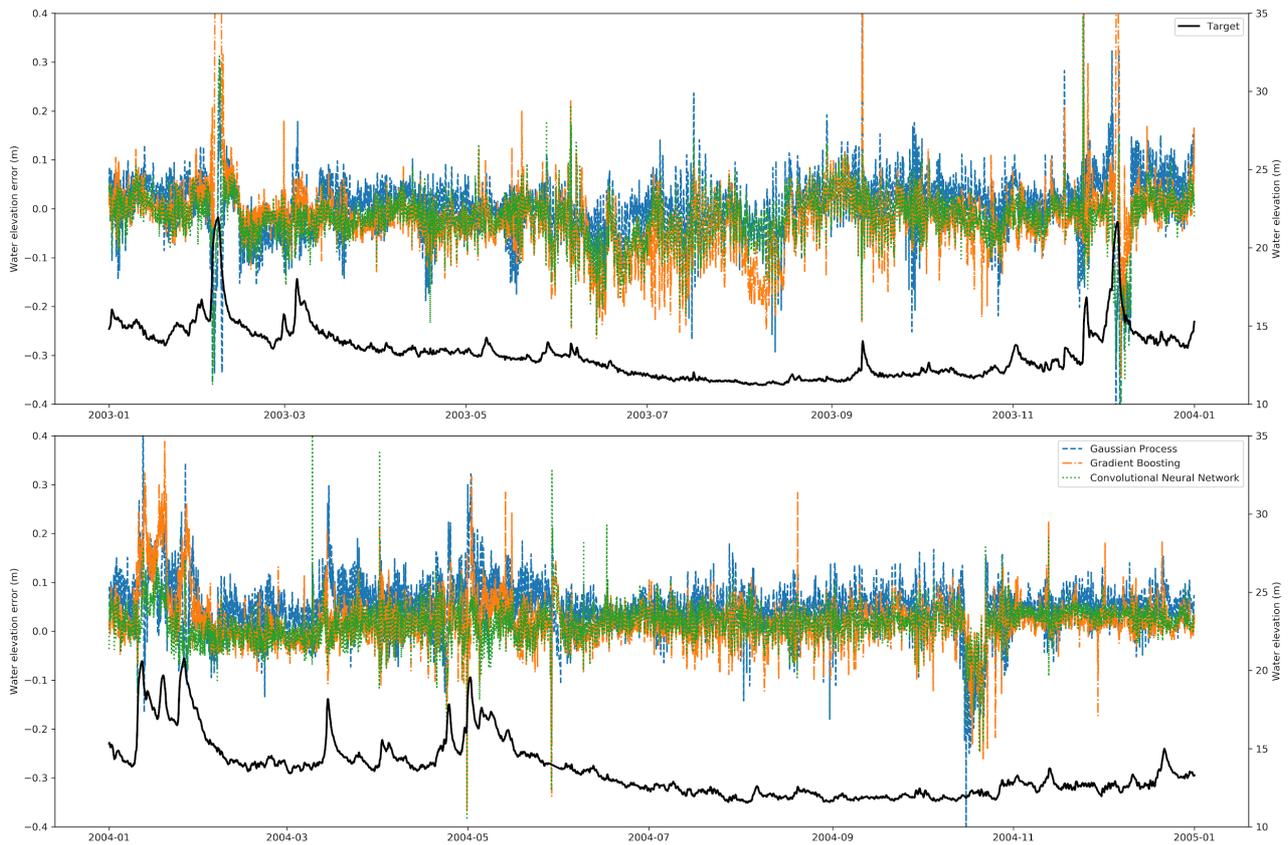

**Figures 6**: Difference between the test target and predicted water level at Marmande for years 2003 and 2004.

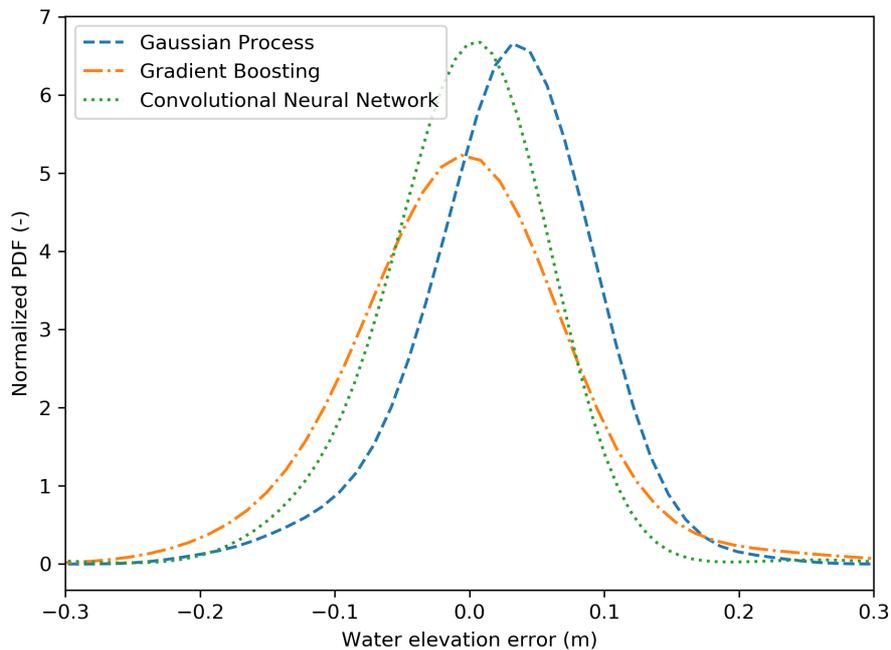

**Figures 7**: Probability Density Function (PDF) of the water elevation error with respect to the target for the test set.

## 4. CONCLUSIONS AND PERSPECTIVES

In this paper, a hydraulic state estimation has been framed as a learning task using boundary condition time series. Various techniques were used: a simple regression, a kriging method, a gradient boosted tree learning approach, and two neural networks. In this process, a baseline was introduced to show how much information was learned beyond the results of the regression. Additionally, the target quantity was taken from 2 sources: a 1D physical model solving the Shallow Water Equations over the period, and the in-situ observations that represent the full physics of the flow. This shed light on merits and shortcomings of the various learning methods. When learning from synthetic data from the physical model, kriging performed best, suggesting that the natural regularity of the data introduced by the solver did not give an edge to machine learning approaches. This contrasts with the in-situ data learning task, on which some machine learning techniques offered a significant improvement on linear regression and kriging. While mean square error remain around the admissible accuracy expected in hydrodynamics modeling (about 5 to 10cm) for all learning models, the maximum error reaches 1m showing that some important caveats appeared with some machine learning approaches. While this metrics was not includes in the learning process, it is important to note that observed extreme events are poorly predicted. This is an important observation, as these events could potentially be critical for population safety, and sheds some light on how the optimization processes use in machine learning must be carefully selected: if an important metric is not directly included at training time, there is little to no guarantee that it will be well met by the training result.

The disruption by ML techniques of more traditional approaches is a highly trending topic, but in the field of physics the uses of ML have yet to be fully determined. While this paper has made some advances in showing how a hydraulic state estimation problem can be framed as a learning one, and some caveats that accompany this strategy, many open questions remain concerning the uses of ML in conjunction with physical models. To start, in future work our focus will shift to hydraulic state forecasting. As this study suggests, exploring more massive databases of in-situ data with learning techniques could yield accurate results, since specifically neural networks are known for their ability to leverage large amounts of data. In a next step, multiple sources of observation data (*e.g.* including satellite observations) could be added to further increase accuracy.

## ACKNOWLEDGEMENTS

The authors gratefully acknowledge the SCHAPI and SPC networks for providing input data on the hydrology of the Garonne river. The authors also acknowledge LNHE for providing the Garonne test case.

## REFERENCES AND CITATIONS

[1] Besnard, A., & Goutal, N. (2011). Comparison between 1D and 2D models for hydraulic modeling of a floodplain: case of Garonne River, *La Houille Blanche*, **3**, 42-47.

[2] El Mocayd, N. (2017). La Décomposition en polynôme du chaos pour l'amélioration de l'assimilation de données ensembliste en hydraulique fluviale. PhD Université de Toulouse, INP.

[3] Gauckler, P. (1867). Etudes Théoriques et Pratiques sur l'Écoulement et le Mouvement des Eaux, *Comptes Rendus de l'Académie des Sciences,* Paris, France, Tome 64, 818–822.

[4] Goutal N., Lacombe J.-M., Zaoui F., & El-Kadi-Abderrezzak K. (2012). MASCARET: a 1-D open-source software for flow hydrodynamic and water quality in open channel networks, River flow 2012, 1169-1174.

[5] Hervouet J.M. (2003). Hydrodynamique des écoulements à surface libre : Modélisation numérique avec la méthode des éléments finis. Presse de l'Ecole Nationale des Ponts et Chaussées.

[6] Friedman, J. H. (2001). Greedy function approximation: a gradient boosting machine. *Annals of statistics*, 1189-1232.

[7] Goodfellow, I., Bengio, Y., & Courville, A. (2016). *Deep learning*. MIT press.

[8] Rosenblatt, Frank (1958), The Perceptron: A Probabilistic Model for Information Storage and Organization in the Brain, Cornell Aeronautical Laboratory, Psychological Review, v65, No. 6, pp. 386–408.

[9] Grossberg, S. (1982). Contour enhancement, short term memory, and constancies in reverberating neural networks. In Studies of mind and brain (pp. 332-378). Springer, Dordrecht.

[10] Rasmussen C., C. Williams (2006). Gaussian processes for machine learning. MIT Press.

[11] Krige D.G., Guarascio M., and Camisani-Calzolari F.A. (1989). Early South African geostatistical techniques in today's perspective. Geostatistics 1, 1–19.

[12] Wales D.J., Doye J.P.K. (1997). Global Optimization by Basin-Hopping and the Lowest Energy Structures of Lennard-Jones Clusters Containing up to 110 Atoms. The Journal of Physical Chemistry A 101(28), 5111–5116.

[13] Pearson K. (1901). On Lines and Planes of Closest Fit to Systems of Points in Space. Philosophical Magazine. 2(11), 559–572.

[14] Wiener, N. (1938). The homogeneous chaos. Am.J.Math 60(4), 897–936

[15] Spanos, P., Ghanem, R. (1991). Stochastic Finite Elements : A Spectral Approach. Springer

[16] Pamphile T. Roy, Nabil El Moçayd, Sophie Ricci, Jean-Christophe Jouhaud, Nicole Goutal, Matthias De Lozzo, and Mélanie C. Rochoux (2017). Comparison of polynomial chaos and Gaussian process surrogates for uncertainty quantification and correlation estimation of spatially distributed open-channel steady flows. Stochastic Environmental Research and Risk Assessment, pages 1–29.

[17] Scikit-Optimize Python library (2019). DOI: 10.5281/zenodo.1157319